\documentclass[11pt]{article}
\usepackage[margin=1in]{geometry}
\usepackage[T1]{fontenc}
\usepackage[utf8]{inputenc}
\usepackage{microtype}
\usepackage{booktabs}
\usepackage{array}
\usepackage{tabularx}
\usepackage{enumitem}
\usepackage{float}
\usepackage{placeins}
\usepackage{tikz}
\usetikzlibrary{arrows.meta,positioning}
\usepackage{hyperref}
\usepackage[round,authoryear]{natbib}
\hypersetup{colorlinks=true,allcolors=blue}
\setlist{nosep}

\title{Beloved Afterlives: Governing AI Resurrection Beyond Consent}
\author{Hanjing Shi \and Dominic DiFranzo\\
Department of Computer Science and Engineering\\
Lehigh University, Bethlehem, PA, USA}
\date{August 2026}

\begin{document}
\maketitle

\begin{abstract}
AI resurrection is often framed as a question of consent: did the represented person authorize being made to speak? That question matters, but it freezes authority at the moment of creation. A representation can later change models, pass to relatives, depend on a provider, incorporate records shared with others, or circulate far beyond its intended audience. We argue that the central governance problem is therefore not whether authorization exists once, but whether it remains legible as the representation moves. Across a public-record audit of 93 systems, creation was far easier to inspect than the conditions for speaking, contesting, preserving, or leaving: consent or authority information was thin in 82 systems, objection or redress in 82, and deletion or export in 77. The differences among systems reveal why these gaps cannot be reduced to one transparency score. Human afterlives show consent becoming incomplete over time. Companion-animal afterlives begin where subject consent is unavailable and shared care must allocate authority. Adjacent persona and mimetic systems show how voices, likenesses, and personalities can travel into later afterlife uses. Four public cases follow the same movement from premortem participation, through intimate postmortem creation, to third-party circulation and family contestation. From this evidence we develop \emph{relational authority}: authorization is distributed across people, records, providers, and audiences, and must remain traceable as those relations change. This reframes AI resurrection from a product authorized once into an accountability chain linking creation authority, source boundaries, circulation, contestation, and exit. The study measures what users and affected parties can inspect publicly; private implementation and lived outcomes remain open empirical questions.
\end{abstract}


\section{Introduction}

A loved one no longer has to be alive, human, or able to consent to become a recurring AI presence. A service can turn stories into a chatbot, recordings into a familiar voice, photographs into a speaking avatar, or shared memories into messages attributed to a deceased companion animal. The public invitation is simple: upload, record, generate, preserve. The difficult questions arrive one step later. Who had authority to make the representation speak? What prevents it from drifting beyond the available record? What can a sibling, former partner, co-caregiver, or estate representative do when they disagree? How does a user leave without losing the underlying materials?

Those questions matter even when nobody mistakes a model for a person. A mourner may know that a reply is generated and still return because it occupies the place of a parent, partner, friend, or animal in an existing relationship. The representation can be both artificial and relationally consequential. This tension shifts the problem beyond deception. It asks how a platform organizes permission, attributed speech, persistence, and exit around a relationship that existed before the system did.

Digital-remains research first showed how platforms mediate the stewardship of a person's records after death \citep{brubaker2014stewarding}. Ghostbot governance scholarship then placed consent, dignity, and competing survivor claims around systems that act in the deceased person's place \citep{harbinja2023governing}. Responsible-griefbot work further distinguished the data donor, data recipient, and service interactant rather than treating ``the user'' as one role \citep{Hollanek_2024}. Generative-ghost research expanded the design space from preservation to novel postmortem interaction \citep{morris2025generativeghosts}. Beyond explicit afterlife products, mimetic AI makes impersonation an infrastructure problem \citep{Bukingolts_2025}. Voice-cloning research shows why identity claims can attach to a generated voice \citep{leuenberger2025the}. Image-rights scholarship traces a parallel claim in digital likeness \citep{Heugas_2021}. This scholarship explains why synthetic presence deserves scrutiny. It does not show, across a product corpus, what a prospective user or affected person can inspect before providing intimate data, money, or consent.

The public layer is consequential because it is often the first and only common evidence available to all parties. The account holder may later see private controls. A represented person, relative, co-caregiver, or other affected party may never receive that access. Landing pages, policies, help centers, pricing pages, and public documentation therefore perform more than a marketing function. They establish what can be known before creation and what can be cited when authority is questioned. A missing public route leaves the outsider's evidence trail incomplete, regardless of whether an internal safeguard exists.

Companion-animal memorial AI makes that endpoint unusually visible. An animal cannot authorize synthetic speech. The photographs, stories, routines, and care memories used to construct that speech may nevertheless belong to several people. Uploader control answers who can press the creation button without settling whose account of the animal becomes canonical, whether another caregiver receives notice, or how shared photographs can be removed without destroying the whole memorial. Human and animal cases share this relational structure while retaining different rights and consent mechanisms. The animal case reveals how much authority remains to be allocated after uploader permission is known.

The argument moves through three levels. Research on digital remains, griefbots, companion AI, and mimetic systems first locates the relational problem: synthetic speech is made from records and relationships whose claims do not disappear when an account is created. A structured audit then asks where that problem is publicly inspectable. Its 93 systems occupy three analytical positions. Fifty-three human systems concern a deceased person and reveal how consent becomes incomplete as a representation persists. Seventeen companion-animal systems carry the authority question into relationships where the represented subject cannot consent, bringing shared care and shared records to the foreground. Twenty-three adjacent systems---17 persona or companion platforms and 6 mimetic-infrastructure services---show how voices, likenesses, and personalities can be captured or generated before they are used for an afterlife. Each position exposes a different break in the same path from creation to circulation and exit.

The four public cases bring relationships back into the audit fields. James Vlahos's Dadbot and Michael Bommer's work with Eternos begin with premortem participation; Joshua Barbeau's use of Project December begins with postmortem intimate creation; the synthetic Qiao Renliang videos carry third-party creation into public circulation and family objection. Counts locate a recurring absence. Cases follow who encounters it as authority changes hands and the representation travels.

We ask three questions:

\begin{enumerate}[label=RQ\arabic*.]
\item What information about creation authority, represented-subject authorization, source grounding, affected-party contestation, deletion, export, and exit is publicly inspectable?
\item How do these public gaps differ across human beloved afterlives, the companion-animal boundary, and adjacent enabling systems, and how do four contrasting cases change their relational meaning?
\item What public controls follow when creation is easier to inspect than authorization, contestation, correction, or exit?
\end{enumerate}

The empirical contribution begins with a denominator and a boundary: a 93-system audit that distinguishes human beloved afterlives, a companion-animal boundary, and adjacent systems that supply representational capacity. The cases then restore the relationships hidden by fields such as ``uploader assertion'' and ``generic report process,'' showing how the same code changes when the creator is the represented person, a partner, a child, or a stranger. From that movement we develop the paper's central concept, \emph{relational authority}: authorization is distributed across people, records, providers, and audiences, and its scope changes as a representation becomes more generative, persistent, or public. Affected-party standing translates that concept into procedure by identifying who may seek notice, review, correction, or selective removal without granting every claimant an equal veto. The resulting accountability chain connects the creation record to source disclosure, circulation checkpoints, contestation, preservation, and exit. Authority must remain legible after the upload.

\section{From Digital Remains to Beloved Afterlives}

\subsection{Postmortem data became interactive presence}

Early digital-afterlife research examined how online memorial spaces extend death and mourning into networked life \citep{brubaker2013beyond}. Work on legacy stewardship showed that survivors and platforms must negotiate access to, and responsibility for, the deceased person's data \citep{brubaker2014stewarding}. A political-economic account of the digital afterlife industry examined the institutions that preserve and monetize those remains \citep{_hman_2017}. Its companion ethical framework translated that structure into obligations among industry actors \citep{_hman_2018}. Across these strands, death does not dissolve relationships among data subjects, survivors, platforms, and audiences. It rearranges their claims.

Generative systems make the rearrangement more active. A static profile can preserve selected records. A conversational bot can respond in the first person. A voice model can deliver words that were never recorded. An avatar can join a new scene or answer a new question. Ghostbot governance research asks who may authorize and control such postmortem action \citep{harbinja2023governing}. Responsible-griefbot scholarship locates obligations among donors, recipients, interactants, and providers \citep{Hollanek_2024}. Work on digital replicas asks what kind of entity, if any, could stand in for a person \citep{Karpus_2025}. Research on interactive deadbots emphasizes the indeterminacy of their technical and social identity \citep{Kozlovski_2026}. Other accounts argue that these constructs can serve purposes beyond grief itself \citep{buben2025beyondgrief}. The central issue is no longer only who stores the data, but who may use those data to produce new attributed acts.

This shift makes source boundaries important. Pre-recorded answers, retrieved excerpts, scripted branches, and open-ended generation do not create the same relation between source and output. A service may preserve a person's own stories, infer a response from a small set of messages, or generate fluent speech with no visible source record. All three may be marketed through continuity. Only the first two offer an obvious path back to what the person actually supplied, and even then the public record may not reveal how retrieval or selection works.

Consent also becomes layered. Premortem participation can authorize collection and an initial representation. It may not settle who inherits control, whether a provider may change the system, whether the representation can enter new domains, or whether relatives can remove shared materials. Postmortem creation raises a harder baseline because a survivor's relationship may support access to source material without establishing broad authority to speak for the deceased. Public-figure cases add another layer. Public availability of images or speech can make technical creation possible while leaving authorization and affected-party claims unresolved.

We use \emph{beloved afterlives} to foreground this relational structure. The term covers AI-mediated representations tied to deceased humans or companion animals and encountered through an existing relationship. Human and animal cases retain different legal rights, consent mechanisms, and governance claims while sharing a narrower transformation: an existing relationship becomes a generative or persistent platform presence. Persona platforms and mimetic tools sit beside this definition as enabling systems because they can capture the voice, likeness, or personality later used in such a presence. Beloved status arises from the prior relationship, not from technical capability alone. The term therefore marks a relation rather than a product family and leaves benefit or harm as an empirical question.

\subsection{Continuing bonds do not settle platform authority}

Bereavement-design research shows that digital belongings can be passed on, maintained, or put to rest through different practices \citep{reference2010passing}. Continuing bonds have been studied among bereaved young people \citep{reference2021continuingbondswithchildrenandbereavedyoung}. The same framework has been applied after companion-animal loss \citep{referencepackman}. Digital memorials may preserve stories, support remembrance, or keep a relationship visible. Analysis of AI grief technologies cautions against building one model of ``normal'' grief into their design \citep{Paumen_2026}. Clinical-thanatology work proposes conditions for responsible integration instead \citep{Manevich_2025}. These findings prevent an easy conclusion that continued interaction is inherently unhealthy.

They also leave a governance problem intact. A continuing bond describes a relation between a living person and a deceased other. A commercial AI product adds a provider, an interface, generated output, subscription terms, moderation, data retention, and technical dependence. The provider does not merely host a memory. It determines what can be uploaded, what the representation can say, what evidence remains, and which person can end the interaction.

This distinction matters for psychological claims. A public-document audit cannot determine whether a representation comforts, unsettles, isolates, or supports a user. It can identify whether support language is paired with crisis resources, age guidance, pause controls, or a clear exit. Those are observable public features. They are not measures of clinical effect.

Companion systems sharpen the same boundary. Long-term chatbot use can develop into a human--chatbot relationship \citep{skjuve2021chatbot}. The loss or deletion of an AI companion can itself become a relational event \citep{Banks_2024}. Synthetic-relationality research connects such intimacy to identity and governance risks \citep{Bhat_2025}. Companion-animal loss is frequently disenfranchised or socially under-recognized \citep{spain2019pet}. Continuing bonds also shape grief after an animal's death \citep{Lykins_2023}. Online memorial spaces can maintain the animal's social presence through images and stories \citep{Eason_2019}. These literatures explain why a pet memorial or pet persona may carry emotional weight. They do not authorize a platform to transform one caregiver's account into the animal's voice without making that transformation inspectable.

\subsection{Public documents are an accountability surface}

Public product materials reveal less than a system audit or ethnography. They also reach a different audience. A prospective user may rely on a product page to decide whether to upload recordings. A relative may rely on a help page to locate a takedown route. A regulator or journalist may rely on a policy page to compare a provider's public commitments. Audits of public privacy policies show how documentation exposes commitments and omissions without verifying implementation \citep{reference2025userprivacyandlargelanguagemodels}. Work on adverse-event reporting specifies the information needed to route and compare reports at scale \citep{Gailmard_2025}. Responsible-reporting research asks how a disclosure reaches actors who can evaluate and act on it \citep{Kolt_2024}.

This paper calls the measured property \emph{public inspectability}. It asks what a person can verify from materials available without creating an account, uploading private data, or entering a support exchange. Public inspectability differs from internal transparency. A provider may operate a strong control that is not publicly documented. The audit would code the public field as thin because the control cannot guide a pre-use decision or an outsider's objection.

Legal and policy instruments offer partial support for these questions. The GDPR creates access, deletion, and complaint mechanisms for personal data \citep{reference2016regulation}. The EU AI Act addresses transparency duties for some synthetic content \citep{union2024regulation}. China's deep-synthesis provisions address the provision and labeling of synthetic media services \citep{china2022provisions}. Its generative-AI measures govern a broader class of services \citep{china2023interim}. Later labeling rules specify marks for AI-generated synthetic content \citep{china2025measures}. California AB 1836 addresses postmortem digital replicas in defined expressive uses \citep{legislature20241836}. California AB 2602 restricts specified contracts for digital replicas \citep{legislature20242602}. Scholarship on China's digital-afterlife law examines how existing doctrines meet AI resurrection \citep{cheng2025thelawofdigitalafterlife}. Analysis of memorial chatbots separately foregrounds privacy and data protection \citep{Ciani_Sciolla_2025}. None begins from the complete social arrangement of a beloved afterlife. We therefore use the legal scan to identify possible control points, not to determine compliance.

\section{Method}

\subsection{Evidence layers and study object}

Each audit record corresponds to one publicly inspectable product or platform. The design distinguishes three analytical positions. Human beloved afterlives and the companion-animal boundary both concern an existing relationship made persistent after death, while remaining separate because consent and legal status differ. Adjacent persona, companion, and mimetic tools have a documented representational use case but no necessary tie to death or a prior intimate relationship. We inspected public landing pages, frequently asked questions, terms, privacy and safety pages, pricing pages, app-store listings, public documentation, product directories, and provider use cases.

The 93 coded systems form the only product denominator. Within it, three analytical positions keep different objects from collapsing into a single category: human beloved afterlives, companion-animal beloved afterlives as a boundary layer, and adjacent enabling systems. These positions organize the audit; they are distinct from the paper's evidentiary progression from scholarship, through corpus patterns, to cases. A targeted set of 91 scholarly sources supplied concepts, prior findings, and candidate products. Sixteen legal and policy documents supplied contextual control points. Neither set was pooled with the product records. This separation prevents a common category error in mixed-source audits: a scholarly discussion or statute can explain why a field matters, but it cannot increase the number of products that publicly disclose the field.

The observed layer ends at the public boundary. We did not create accounts, upload memorial data, purchase services, test generated replies, submit takedown requests, inspect moderation queues, or interview users and providers. We cannot determine whether a private control exists, whether a deletion request succeeds, whether a model stays faithful to source material, or how use affects grief.

\subsection{Retrieval, inclusion, and corpus closure}

Product discovery used fixed query families for AI afterlife, griefbot, deadbot, digital legacy, memorial chatbot, memorial voice clone, posthumous avatar, companion-animal memorial AI, persona clone, skills clone, digital twin, and memorial media generation. Searches covered general web results, official pages, public app-store and product-directory entries, vendor use cases, products named in scholarship, and citation leads. Targeted long-tail searches sought memorial-photo, memorial-video, pet-tribute, pet-chat, and generated-voice products that broad afterlife queries could miss.

A system entered the core stratum when public documentation described creating, hosting, preserving, reconstructing, or simulating a deceased human or companion animal. A living-person persona or companion platform entered the adjacent stratum when its public materials described capturing or generating a particular person's identity, voice, likeness, or personality. A generic image, voice, or video tool entered that stratum only when the provider published a memorial, deceased-loved-one, pet-loss, afterlife, persona, or mimetic use case. Ordinary funeral pages, purely physical memorial services, login-gated flows without an inspectable public description, and news-only mentions without a public product artifact were excluded.

The final corpus contains 29 human afterlife services, 24 static human memorial systems, 17 companion-animal memorial or AI services, 17 persona or companion platforms, and 6 mimetic-infrastructure services. Corpus closure was procedural. After the fixed query families, app-store and directory checks, literature leads, long-tail memorial and pet searches, and a boundary-validation pass dated April 30, 2026, additional candidates either duplicated an existing system, lacked inspectable product material, or failed the inclusion rule. Closure bounds this retrieved corpus. It does not make it representative of an international market. Table~\ref{tab:positions} pairs each analytical position with the question it carries into the audit.

\begin{table}[t]
\centering
\caption{Three analytical positions within the 93-system public product corpus.}
\label{tab:positions}
\small
\begin{tabularx}{\textwidth}{>{\raggedright\arraybackslash}p{0.23\textwidth}>{\raggedright\arraybackslash}p{0.28\textwidth}r>{\raggedright\arraybackslash}X}
\toprule
Analytical position & Included product families & Systems & Question carried forward \\
\midrule
Human beloved afterlives & Human afterlife (29); static human memorial (24) & 53 & How does consent change as a representation persists, changes, or passes to others? \\
Companion-animal boundary & Pet memorial or AI (17) & 17 & Who holds authority when the subject cannot consent and care memories are shared? \\
Adjacent enabling systems & Persona or companion (17); mimetic infrastructure (6) & 23 & How do voice, likeness, and persona capabilities travel into later representational uses? \\
\midrule
Full denominator & Five product families & 93 & Where does the public evidence chain break? \\
\bottomrule
\end{tabularx}
\end{table}

Retrieval relied mainly on English search terms and English-accessible pages. Some non-English legal and case materials were included where they could be verified, but the product corpus cannot characterize markets in China, South Korea, Japan, or other regions. The corpus also overrepresents products with stable public pages. Small, discontinued, private, or app-only services are easier to miss.

\subsection{Coding and decision rules}

One primary coder applied a codebook covering product family, represented subject, system mode, interactivity, consent model, authority model, data source, represented-person verification, source grounding, synthetic labeling, objection and redress, deletion, export, co-caregiver or family conflict, support framing, crisis resources, age or life-stage language, pricing pressure, emotional monetization signals, and confidence. Each system record retains inspected URLs, access information, selected codes, an evidence paraphrase, and notes.

The coding separated several features that public pages often blend. Represented-person verification asked whether the subject participated before death, completed an identity process, entered through an enterprise contract, depended on uploader assertion, or remained unclear. Source grounding distinguished source-grounded output, scripts or recorded answers, general model guardrails, unclear boundaries, and no visible boundary. Synthetic labeling was coded separately because a clear AI label can coexist with weak source grounding.

Redress coding also distinguished routes. A clear impersonation or takedown process offered more public guidance than a generic report button. A contact email showed a point of contact but not a review workflow. Deletion and export were separate because removal without export may force a user to choose between continued platform dependence and loss of source materials. Shared-care conflict was coded only when public materials addressed disagreement among relatives or caregivers, rather than merely allowing collaborative uploads.

We use \emph{thin public information} for absent, ambiguous, generic, or requester-dependent material under the submitted coding rules. The term does not mean that no internal control exists. It means a public reader cannot reconstruct the applicable process with confidence.

Twenty systems were marked low confidence because public access was limited, documentation was thin, or the family boundary was unstable. Removing those systems preserved the direction of the main findings. Among the remaining 73, consent or authority information was thin in 64, objection or redress in 62, deletion or export in 58, grounding in 34, and mental-health safeguards in 32. The analysis used one primary coder and does not establish intercoder reliability.

\subsection{Four-case public-record comparison}

The case comparison asks what the aggregate audit cannot. Counts show that a field is thin. Cases show which relationship must carry the missing authority and how public circulation changes the dispute. We selected Dadbot, Michael Bommer and Eternos, Joshua Barbeau and Project December, and Qiao Renliang synthetic videos because they vary on premortem participation, creator relationship, provider involvement, generative scope, circulation, and visible objection.

The evidence follows authority through four connected movements. Dadbot and Eternos begin with premortem participation and carry authorization forward into questions of inheritance, model change, and recipient choice. Project December moves to intimate postmortem access, where relational knowledge enables reconstruction while the represented person can no longer delimit new speech. The Qiao videos carry third-party synthesis into public circulation, adding relatives, platforms, audiences, and removal demands. The 17 companion-animal systems press the same analysis past individual consent: when the represented subject cannot authorize speech, shared records and caregiving histories become the ground on which authority must be allocated.

\begin{figure}[H]
\centering
\begin{tikzpicture}[
  scale=0.94,
  transform shape,
  font=\small,
  evidence/.style={draw, rounded corners=2pt, align=center, text width=4.55cm, minimum height=1.25cm, fill=black!5, inner sep=4pt},
  stage/.style={draw, rounded corners=2pt, align=center, text width=2.62cm, minimum height=1.35cm, fill=white, line width=0.7pt, inner sep=4pt},
  case/.style={align=center, text width=2.8cm, font=\footnotesize, inner sep=2pt},
  flow/.style={-{Latex[length=2.2mm]}, line width=0.8pt},
  evidenceflow/.style={-{Latex[length=1.8mm]}, line width=0.6pt, black!65}
]
\node[evidence] (human) at (0,4.25) {Human afterlives ($n=53$)\\Consent changes as a representation persists};
\node[evidence] (animal) at (5.25,4.25) {Companion animals ($n=17$)\\Shared care allocates authority when the subject cannot consent};
\node[evidence] (adjacent) at (10.5,4.25) {Adjacent systems ($n=23$)\\Voice, likeness, and persona capacity travel across contexts};

\node[stage] (create) at (-0.45,1.8) {\textbf{Create}\\authority record};
\node[stage] (generate) at (2.95,1.8) {\textbf{Generate}\\source boundary};
\node[stage] (circulate) at (6.35,1.8) {\textbf{Circulate}\\authorization and provenance};
\node[stage] (contest) at (9.75,1.8) {\textbf{Contest}\\affected-party review};
\node[stage] (exit) at (13.15,1.8) {\textbf{Exit}\\export, pause, transfer, deletion};

\draw[flow] (create) -- (generate);
\draw[flow] (generate) -- (circulate);
\draw[flow] (circulate) -- (contest);
\draw[flow] (contest) -- (exit);

\draw[evidenceflow] (human.south) -- ++(0,-0.35) -| (create.north);
\draw[evidenceflow] (animal.south) -- ++(0,-0.55) -| (contest.north);
\draw[evidenceflow] (adjacent.south) -- ++(0,-0.35) -| (circulate.north);

\node[case] (premortem) at (-0.45,0.15) {Dadbot and Eternos\\premortem participation};
\node[case] (intimate) at (2.95,0.15) {Project December\\intimate postmortem creation};
\node[case, text width=5.4cm] (public) at (8.05,0.15) {Qiao Renliang videos\\third-party public circulation and family contestation};
\draw[evidenceflow] (premortem.north) -- (create.south);
\draw[evidenceflow] (intimate.north) -- (generate.south);
\draw[evidenceflow] (public.north) -- (circulate.south);
\end{tikzpicture}
\caption{The paper's evidence chain. The three audit positions locate different breaks as a representation moves from creation to exit; the four cases show how authority changes with participation, generative extrapolation, and public circulation.}
\label{fig:evidence-chain}
\end{figure}
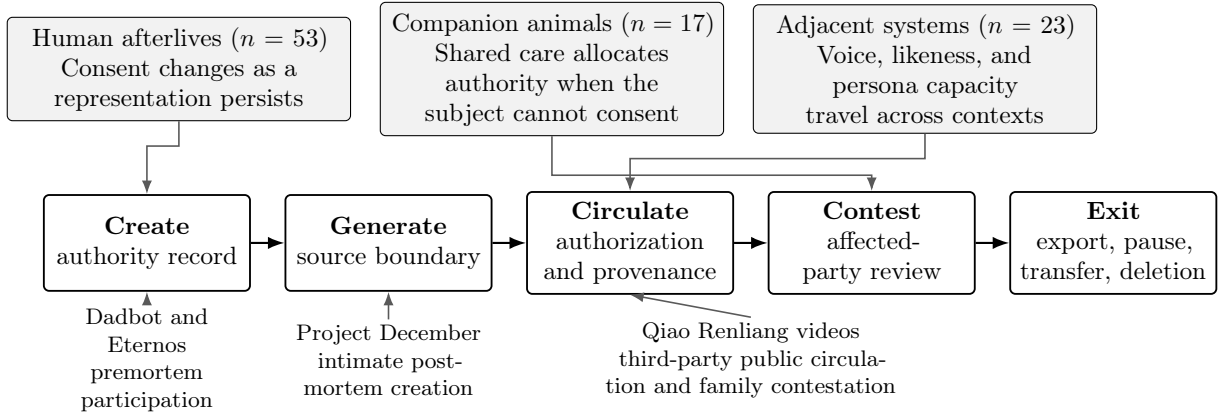

For each case, we recorded the creator, represented person, participation before death, source material, representation mode, grounding description, circulation, affected parties, contestation, response, persistence, and exit. A central chronology required two independent public sources or one direct source for a narrower system-description claim. Provider statements were labeled as provider claims. Unknown fields remained unknown. We did not count passages or convert the four cases into percentages.

The cases remain limited. Public reporting can omit private family discussions, platform communications, later deletion, and technical change. The Qiao materials are especially dependent on reporting that quotes family statements and describes platform content after public circulation. We therefore distinguish a reported request for removal from a verified platform-wide outcome.

\section{Findings from the 93-System Audit}

\subsection{The three analytical positions reveal different pressure points}

Grouping the two human product families yields 53 systems. Consent or authority information was thin in 51, objection or redress in 49, and deletion or export in 45. These systems begin with a deceased subject, an uploader, and records embedded in an existing relationship. Their most persistent opacity concerns who may create, who may challenge, and how initial consent changes as the representation persists, moves to another person, or depends on a provider.

The 17 companion-animal systems intensify one part of that pattern: all were thin on consent or authority, and 15 were thin on objection or redress. By removing the assumption that the represented subject could have authorized speech, they isolate uploader power, shared records, and caregiving relationships as separate sources of authority.

The 23 adjacent systems look different again. Source grounding was thin in 19 and deletion or export in 20. Their significance lies in the portable representational capacity they assemble. A voice, persona, or likeness may be captured while someone is alive, reused across providers, and later placed into a postmortem setting. The adjacent layer locates that upstream dependence on inference and provider infrastructure before it enters an afterlife relationship.

\begin{table}[H]
\centering
\caption{Selected public-information gaps by analytical position and across the full corpus.}
\label{tab:gaps}
\begin{tabular}{lrrrr}
\toprule
Field coded as thin & Human ($n=53$) & Animal ($n=17$) & Adjacent ($n=23$) & All ($N=93$) \\
\midrule
Consent or authority & 51 & 17 & 14 & 82 \\
Objection or redress & 49 & 15 & 18 & 82 \\
Deletion or export & 45 & 12 & 20 & 77 \\
Source grounding & 19 & 6 & 19 & 44 \\
\bottomrule
\end{tabular}
\end{table}

\subsection{Creation authority is visible before refusal authority}

Fourteen systems foregrounded premortem self-recording or self-creation. Eleven described identity verification, and three relied on an enterprise contract. By contrast, 53 relied on uploader assertion and 12 left represented-person verification unclear. Thus 65 of 93 made represented-person verification thin at the public boundary. When family or co-caregiver conflict visibility was included, consent or authority information was thin in 82 systems.

The recurring pattern was procedural imbalance. Public pages gave a prospective creator an evident path: begin an interview, upload photographs, add messages, select a voice, describe a personality, or generate a tribute. A sibling, former partner, estate representative, or co-caregiver rarely received an equally specific path to question that authority before publication. Creation appeared as a workflow. Refusal appeared, if at all, as a general contact channel.

Premortem participation improved one part of the record. A person who recorded stories or configured a future representation supplied visible evidence of creation authority. That evidence did not automatically answer whether later model changes were within scope, who inherited control, whether family members could remove shared media, or what happened when the provider closed. The audit therefore separates initial authorization from continuing authority.

Static memorial systems did not escape the issue. All 24 static human memorial systems were coded thin on consent or authority under the family-level gap rule. Their lower interactivity reduces the risk of open-ended first-person speech, but a static page can still use disputed images, biographies, or relational records. The authority problem narrows. It does not disappear.

\subsection{Presence claims outpace source grounding}

Forty-five systems were script- or recording-bounded. Four were coded source grounded. Seventeen referred to model guardrails without making the source relationship clear, 16 were unclear, and 11 showed no visible grounding policy. Under the submitted gap rule, 44 of 93 systems had thin public grounding information.

This result matters most when representation becomes generative. A pre-recorded answer can be traced back to the speaker. A retrieved passage can be compared with a source collection. A voice clone can deliver new sentences while sounding familiar. An open-ended persona may produce plausible first-person claims without showing which memory, message, interview, or public record supports them. Public pages often emphasized voice, personality, presence, or realism before explaining this distinction.

Synthetic labeling solved a different problem. Thirty-four systems used a clear label, 32 relied on contextual labeling, 25 were unclear, and 2 did not require a synthetic label under the coding rule. A label can tell a user that content is generated. It cannot tell the user whether the content repeats a recording, retrieves a documented memory, or invents a likely answer. Public controls need both identity disclosure and source disclosure.

The audit does not show that weakly documented systems hallucinate more or that users believe them. It identifies the missing evidence required to assess those possibilities. A source list, output-level citation, statement of generative scope, or clear boundary between recorded and inferred speech would make the representation easier to inspect before reliance.

\subsection{Contestation and exit lag behind creation}

Objection or redress information was thin in 82 systems. Fifty-two pointed to a generic report process, 12 offered only a contact email, 18 showed no visible route, and 11 described a clearer impersonation or takedown process. The difference is substantial. A generic form may accept a complaint. It may not tell a non-account-holder what evidence to provide, whether disputed content will be paused, who reviews competing claims, or how a decision can be appealed.

Deletion and export showed a related asymmetry. Fifty-three systems described request-based deletion, 16 account deletion, 15 left deletion unclear, and 9 showed no visible deletion control. Only 15 systems offered clear export; 53 offered partial export and 25 showed none. Under the combined coding rule, deletion or export was thin in 77 systems.

These controls serve different moments. Export preserves stories, recordings, or photographs before a provider relationship ends. Deletion removes an account or representation. Billing termination stops commercial continuity. Transfer moves control to another person. Objection allows someone without account ownership to question the artifact. A page that says ``contact us to delete your data'' does not explain all four.

The asymmetry is most consequential when a representation depends on the provider. A static file can often be stored elsewhere. A hosted conversational voice may require proprietary retrieval, synthesis, moderation, and account infrastructure. Exit can therefore mean losing the interaction even when a user keeps the underlying recordings. Public pages rarely distinguished these layers.

\subsection{Companion-animal systems reveal distributed authority}

None of the 17 companion-animal systems described how disagreement among caregivers would be handled. The subgroup included static tributes, generated images and videos, AI-written memorial text, saved voice, chat, and animal-persona features. Some outputs were clearly framed as generated or imaginative. The authority question remained because the input materials and relationship were often shared.

Consider a simple conflict. One person uploads photographs and routines from a jointly cared-for dog, then configures messages in the dog's voice. Another caregiver accepts the memorial photographs but rejects the synthetic speech. Account ownership gives the uploader technical control. It does not resolve whether shared media may be separated from the generated persona or whether the second caregiver should receive notice.

The animal cannot supply premortem consent in the human sense, correct a false memory, or object to attributed speech. This makes uploader consent structurally incomplete. The relevant public record should instead identify the uploader's relationship, disclose that animal speech is generated, distinguish shared media from generated output, and offer a route for another caregiver to request review. The evidence supports this procedural implication. It does not show how caregivers would use the route or whether it would improve mourning.

\subsection{Support language is easier to find than escalation guidance}

Thirty-five systems used comfort or support framing, 22 used healing or closure language, and 6 made clinical-style claims. Twenty-seven stayed with memory language, while 3 used no mental-health framing. Among the support-framed systems, 40 did not pair those claims with an explicit crisis resource under the submitted coding rule. Age or life-stage information was thin in 17 systems.

The finding is modest but useful. Public pages can invoke emotionally serious outcomes while leaving escalation and exit less visible. We cannot infer that users in crisis accessed these products, that the products worsened distress, or that minors used them without protection. We can ask for consistent public architecture: support claims should sit near age guidance, crisis resources, pause options, and a clear statement that the service does not replace professional care where applicable.

Across the full corpus, represented-person verification depended on uploader assertion or remained unclear in 65 systems. None of the 17 companion-animal systems described a process for disagreement among caregivers. These findings add two procedural details to the four cross-stratum gaps in Table~\ref{tab:gaps}: the public record often stops before either the represented subject or another contributor to the relationship can be heard.

\section{Four Public Cases}

\subsection{Dadbot: participation can bound a representation}

James Vlahos began with an oral-history project after his father, John Vlahos, received a terminal diagnosis \citep{vlahos2017dadbot}. An NPR interview describes about twelve hour-long interviews and professional transcription of roughly 200 pages \citep{npr2017dadbot}. Vlahos then used a pattern-matching system to connect user prompts with authored branches, stories, jokes, songs, and edited recordings \citep{abc2017dadbot}.

Authority was unusually legible. John Vlahos participated while alive and supplied much of the source material. The son's role was creative and technical, but the father's recorded words remained visible within the construction. Grounding was also comparatively inspectable because the bot's most characteristic outputs came from interviews and audio rather than unrestricted model generation.

The case still contains limits. Premortem participation did not appear as a formal scope document in the public materials we reviewed. The record does not establish a general family process for revocation, deletion, inheritance, or disagreement. Access remained largely private and creator-controlled, which reduced public circulation but concentrated persistence in the builder. Dadbot therefore separates visible origin from durable governance: participation clarified who spoke and from which archive, while revocation, inheritance, and continuity remained personal arrangements rather than a public process.

Dadbot also marks an important technical boundary. The system could fail to answer or route a prompt awkwardly, but its bounded structure made some absences visible. Open-ended generation can feel smoother while obscuring which statements exceed the record. The case suggests that a less fluent representation may sometimes be more accountable because its limits can be inspected.

\subsection{Michael Bommer and Eternos: self-authorization enters a platform}

Michael Bommer worked with Eternos after a terminal cancer diagnosis to create an interactive voice-based representation. Associated Press reporting describes direct participation, the recording of 300 phrases for voice synthesis, and autobiographical answers about his life and views \citep{grieshaber2024mourners}. Eternos presented the system as able to answer questions and tell stories without simply replaying fixed recordings \citep{eternos2024michael}.

This case begins with strong creation authority. Bommer participated, supplied material, and discussed the project publicly. It also introduces greater platform dependence than Dadbot. The representation used a commercial provider, voice synthesis, an internal model, and external large language models according to the provider account reported by the Associated Press. A user could therefore hear a familiar voice producing new responses whose relation to the recorded archive was not fully visible.

Public materials described private family access and provider claims that rights or control could pass to relatives. A German report stated that the representation would belong to Bommer's wife and that she could decide whether to delete it \citep{faz2024bommer}. These are meaningful claims. They remain partly dependent on provider description and reporting rather than an inspected inheritance workflow or tested deletion process.

The case also prevents a simple story about user demand. Bommer welcomed the project. His wife expressed uncertainty about whether she would use it after his death. That difference does not establish conflict. It shows why one person's authorization cannot predict another person's relationship with the resulting system. A public control should preserve the recipient's ability not to activate, to pause, or to delete without framing non-use as a failure to remember.

Eternos therefore moves the problem forward. Self-authorization can be clear while continuing authority remains distributed among the represented person, relatives, and provider. The more capable the system becomes, the more important it is to state which outputs come from recorded memory, which are inferred, who may extend the source set, and what persists when the provider or model changes.

\subsection{Project December: intimacy supplies authority that the deceased did not document}

Joshua Barbeau used Project December to configure a text simulation of his deceased fiancee, Jessica Pereira. Contemporary reporting describes the old texts, Facebook messages, short description, and underlying generative language model used in the simulation \citep{fagone2021jessica}. Barbeau's direct account confirms that the exchanges continued beyond retrieval of fixed messages and produced new text in a simulated Jessica voice \citep{barbeau2021ama}.

The creator's relationship supplied the practical authority. Barbeau possessed messages, memories, and intimate knowledge. That access made the reconstruction possible. The public record we reviewed did not establish premortem authorization from Pereira for this use. The distinction matters because intimacy can support fidelity judgments without resolving whether a partner may create new attributed speech after death.

Grounding was thinner than in Dadbot. A limited set of messages and a short prompt supported generative extrapolation. The system could produce emotionally coherent statements that were not drawn directly from Pereira's records. Public accounts make this generative scope central to the case. They do not provide an output-level source record through which another person could distinguish remembered phrase, prompt instruction, and model invention.

The platform also shaped persistence. Access depended on a third-party service and model context rather than a locally bounded archive alone. The public case record does not establish a durable correction process, shared-family review, or export of the interactive system. Nor does it document a formal objection from another affected party. These absences should not be converted into assumptions about family consent. They mark what the public record cannot answer.

Project December shows why postmortem intimate creation is ethically difficult even when the creator acts privately and with care. The partner may be the person best able to recognize style and significance. The represented person is still unable to approve the new domain of speech, correct extrapolation, or limit circulation later. The case exposes an authority gap rather than proving misuse.

\subsection{Qiao Renliang videos: public circulation produces contestation}

In March 2024, synthetic videos of several deceased Chinese celebrities circulated on short-video platforms. Red Star News described a generated Qiao Renliang greeting fans and reported his father's statement that the family had not been consulted and wanted the video removed \citep{redstar2024qiao}. A Legal Daily investigation connected such celebrity recreations to a broader commercial market, including paid training in AI recreation \citep{zhao2024resurrection}.

This case differs sharply from the other three. The represented person did not participate. The creator was not presented as an intimate partner or family member. The artifact circulated publicly, so the affected group expanded beyond a private account holder to relatives, fans, platforms, and commercial audiences. Technical access to public images or recordings became a public claim to make the person speak.

Contestation was visible because the family objected. That visibility does not make the remedy complete. Reports described requests for removal and later claims that content had been taken down. The retained public record does not let us verify a platform-wide removal, preservation of evidence, re-upload prevention, or a formal appeals process. We therefore treat the family's objection and the reported response as separate events.

The case shows why public circulation changes authority. A private creator may decide whether to open an app. A publicly circulated avatar enters other people's feeds and can use a person's likeness to attract attention or sell a service. Family objection is not identical to ownership of all public memory; it demonstrates that circulation reaches people who did not create the artifact and who need a route to challenge the claimed authority.

The Qiao case also makes source disclosure insufficient by itself. A page could label the video as AI-generated and still leave authorization unresolved. Conversely, family permission would not establish that the synthetic words were source grounded. Public accountability requires at least three distinct records: who authorized the representation, what sources and inference produced it, and what route exists for objection and removal.

\subsection{Cross-case finding: authority changes with circulation}

The four cases do not form a scale from good to bad. They reveal different breaks in the action chain. Dadbot had direct participation and bounded sources but little formal continuity governance. Eternos also had direct participation and explicit provider involvement, although grounding and long-term control depended more heavily on platform claims. Project December used an intimate partner's records to support generative extrapolation without documented premortem authorization. The Qiao videos used third-party creation and public circulation, producing an explicit family objection.

Two patterns follow. First, premortem participation clarifies creation authority and source origin, but it does not settle inheritance, model change, recipient choice, or provider shutdown. Second, circulation creates affected parties. A private memorial and a viral video may use similar synthesis tools while requiring different notice, contestation, and evidence-preservation procedures.

The cases also show why the 93-system counts need a relational interpretation. A field marked ``uploader assertion'' does not identify whether the uploader is a child recording a consenting parent, a fiance using private messages after death, or a stranger animating a public figure. The same code can conceal very different authority bases. The case layer restores those relationships without turning four cases into prevalence evidence. Table~\ref{tab:cases} condenses that movement without ranking the cases on a single ethical scale.

\begin{table}[t]
\centering
\caption{Cross-case comparison of how authority changes with participation, generation, and circulation.}
\label{tab:cases}
\small
\begin{tabularx}{\textwidth}{>{\raggedright\arraybackslash}p{0.15\textwidth}*{4}{>{\raggedright\arraybackslash}X}}
\toprule
Dimension & Dadbot & Eternos & Project December & Qiao videos \\
\midrule
Participation & Direct premortem interviews & Direct premortem recording & None documented & None documented \\
Creator relation & Son and father & Self with provider & Intimate partner & Third-party creator \\
Mode & Bounded text and audio & Generative voice & Generative text & Synthetic public video \\
Circulation & Private or family & Private family access claimed & Private use reported & Public platform circulation \\
Visible objection & None documented & Recipient uncertainty, not objection & None documented & Family request for removal \\
Central gap & Formal continuity & Grounding and provider dependence & Authorization and extrapolation & Authorization, removal, replication \\
\bottomrule
\end{tabularx}
\end{table}

\section{Keeping Authority Legible from Creation to Exit}

Relational authority becomes governable only when it leaves a record at each transition. The audit locates where that record thins; the cases show how the missing link changes a relationship. The following controls rebuild the chain at the same public boundary measured by the study. Their effects remain questions for evaluation.

\subsection{Record authority before creation}

Every representation should carry a public or affected-party-accessible authority record. The record should identify the creator's role, whether the represented person participated, what right or relationship supports the upload, and whether another person may seek review. It should distinguish premortem self-authorization from family assertion, intimate-partner creation, enterprise contracting, and third-party public-figure use.

The record need not expose private family information to the world. It must be available to people asked to trust or contest the representation. A simple statement such as ``created from interviews recorded by the represented person'' provides a different basis than ``created by a family member from uploaded messages.'' The distinction allows a platform to route disputes without pretending that all uploads carry equal authority.

Affected-party standing should be procedural. A person has a plausible claim to notice or review when the representation incorporates their data or likeness, shared relational records, caregiving history, family or estate role, or direct exposure to the artifact. Standing does not give every affected person an automatic veto. It requires the provider to accept an objection, preserve the contested material, identify the current authority basis, and explain the review outcome.

\subsection{Separate recorded, retrieved, and inferred speech}

Public pages should state whether outputs replay recordings, retrieve source passages, follow scripted branches, or generate new statements. Output-level cues should preserve this distinction during use. A familiar voice should not erase the difference between something the person said and something a model inferred.

For bounded systems, the source record can be simple: interview date, topic, and recording. For retrieval systems, citations can point to a story or message collection. For open-ended generation, the interface should disclose that the response may extend beyond the record and should offer a source view when one exists. These controls support inspection. They do not guarantee truth or emotional safety.

Mode matters. Static memorials need provenance for text and images. Voice systems need voice authorization and a boundary on novel speech. Avatars need likeness and scene provenance. Conversational systems need source traces, inference disclosure, and correction. Public videos need durable synthetic labels and a takedown route that remains usable by non-account-holders.

\subsection{Make objection a workflow}

A complaint channel becomes contestation only when it connects a claim to review and possible remedy. Public documentation should say who may object, what evidence is needed, whether the representation will be paused, how the creator responds, who makes the decision, and whether an appeal exists. The process should preserve a record before removal so that neither the creator nor objector loses the basis of the dispute.

Different remedies fit different conflicts. A disputed photograph may be removed selectively. A generated voice may be paused while authorization is checked. A co-caregiver may request separation of shared media from synthetic speech. A public-figure video may require removal and re-upload controls. Automatic deletion of the entire memorial can harm the account holder, while unrestricted persistence can harm the objector. A review process makes narrower remedies possible.

\subsection{Treat export, deletion, transfer, and exit as distinct}

Users should know what can be exported, in what format, and whether the export preserves only source files or the interactive structure. Deletion documentation should identify timing, backups, public copies, and what happens to generated derivatives. Transfer documentation should state who can inherit access and whether recipients can decline.

Exit has an interactional dimension. A person may want to pause prompts, silence a voice, stop scheduled messages, cancel billing, or archive the representation without deleting the source materials. These options are not equivalent. Providing only account deletion can make disengagement unnecessarily final, while providing only pause can leave data and billing unresolved.

Provider continuity also deserves a public answer. If the service closes or changes its underlying model, can users retrieve recordings, transcripts, images, and configuration? Can the representation be migrated? A platform cannot guarantee indefinite operation. It can make dependence visible before people commit irreplaceable materials.

\subsection{Design for shared care}

Companion-animal systems need a shared-care protocol. At creation, the service should record the uploader's relationship and allow optional identification of other caregivers. It should separate source media, memorial text, and generated animal speech so that a dispute about one layer does not automatically erase the others. Another caregiver should be able to request review without pretending to speak for the animal.

The same architecture can help human families. Shared photographs, messages, and stories often include more than the represented person. Selective review respects that relational provenance. It also prevents the represented subject's consent from being stretched to cover every contributor's data.

Figure~\ref{fig:evidence-chain} condenses these controls into one sequence. Its value is cumulative: an authority record without a source boundary cannot show what the subject actually supplied; provenance without a circulation checkpoint can be detached from the artifact; a complaint route without preservation and remedy does not produce contestation; deletion without export can turn exit into loss. Shared-care review and support guidance enter at the stage where their underlying claim arises rather than appearing as stand-alone disclosures.

\FloatBarrier
\section{Discussion}

AI resurrection is commonly governed as if authorization were a property of the upload: identify who created the artifact, determine whether consent existed, and treat the question as settled. The audit and cases support a different account. Authority changes when a representation begins to generate novel speech, when control passes to another person or provider, when shared records enter the model, and when the artifact reaches people who never chose it. \emph{Relational authority} names this moving distribution of claims over who may create, shape, circulate, contest, preserve, or end a representation.

The object of governance is authority moving across time. Consent records an initial relation between a person and a proposed use; it cannot by itself describe later model changes, inherited control, public recirculation, or claims grounded in shared memory. Public documentation therefore has to preserve an evidence chain across time. The central question is not only who authorized the first act of resurrection, but whether the representation remains answerable as its sources, custodians, and audiences change.

\subsection{Public inspectability is where authority becomes governable}

Legacy-stewardship research describes postmortem data as a negotiation among the deceased, survivors, and platforms rather than an ordinary transfer of files \citep{brubaker2014stewarding}. Audits of frontier developers' privacy policies similarly show that public documents can reveal consequential defaults and omissions even when they cannot verify internal practice \citep{reference2025userprivacyandlargelanguagemodels}. Work on responsible reporting argues that reliable information must reach actors capable of evaluating and acting on emerging risks \citep{Kolt_2024}. Adverse-event scholarship pushes that premise into post-deployment monitoring, where structured reports make evolving failures visible over time \citep{Gailmard_2025}.

We extend this line of work from reporting about a system to answering for a representation. A beloved afterlife does not produce one fixed disclosure problem. Before creation, an outsider may need evidence of the creator's authority. During interaction, a recipient needs to know whether a reply was recorded, retrieved, or inferred. After circulation, a relative or co-caregiver needs a route that connects objection to review and remedy. At exit, the account holder needs to know whether source materials can be preserved without preserving the synthetic presence. These questions form a sequence because each stage inherits unresolved claims from the one before it.

The stratified audit shows where that sequence breaks. Consent or authority was thin in 51 of 53 human systems, while objection or redress was thin in 49. All 17 companion-animal systems were thin on consent or authority, and none described how disagreement among caregivers would be handled. Adjacent systems concentrated opacity elsewhere: source grounding was thin in 19 of 23 and deletion or export in 20. One aggregate transparency score would hide this movement from uncertain creation authority, through uncertain attributed speech, to uncertain exit.

Transparency is therefore too coarse when it is treated as a quantity of disclosure. A synthetic label can identify fabrication without identifying who authorized the artifact. A privacy policy can describe deletion without showing whether a family can preserve irreplaceable recordings. A report button can collect a complaint without giving the complainant standing, an interim hold, or a reasoned outcome. We argue that public inspectability should be judged by whether a person can follow authority from claim to evidence to remedy. When the trail ends at ``contact us,'' the governance problem is not simply missing information; it is that nobody outside the account can see how authority would be tested.

\subsection{Consent begins authority; it does not finish it}

Hollanek and Nowaczyk-Basińska argue that responsible re-creation requires attention to the data donor, data recipient, and service interactant, including mutual consent between donor and interactant \citep{Hollanek_2024}. Harbinja and colleagues foreground the deceased person's wishes and propose an enforceable ``do not bot me'' clause \citep{harbinja2023governing}. Haneman's legal analysis shows that digital resurrection also crosses data, likeness, contract, property, and postmortem personality doctrines \citep{haneman2025the}. These interventions make consent more demanding and more role-aware than uploader permission.

Our cases expose the temporal limit that remains. John Vlahos's participation made Dadbot's origin and source record unusually legible, yet it did not establish a public rule for inherited control or later modification. Michael Bommer authorized his Eternos representation, while its future operation depended on a spouse, provider, and models that could change after his death. Jessica Pereira left no documented authorization for Project December, although Joshua Barbeau's intimacy and possession of messages gave him the practical ability to build it. Qiao Renliang's family became visible as a claimant only after a third party had placed synthetic speech into public circulation. Consent was absent in some cases and unusually strong in others. In none did it govern the representation's entire afterlife.

The companion-animal layer takes the argument one step further. Research on pet loss documents how companion-animal grief is often socially disenfranchised \citep{spain2019pet}. Continuing bonds can remain important after an animal's death \citep{Lykins_2023}. Online memorial spaces can sustain the animal's social presence through shared images and stories \citep{Eason_2019}. These relationships are meaningful without creating a human-like right of animal consent. They instead reveal that uploader permission is already an allocation of authority among caregivers, photographs, routines, and memories. Because none of the 17 systems made that allocation contestable, account control became the default answer to a shared-care question.

Relational authority supplements consent by carrying it forward and placing it beside other grounds for a claim. A person's own image may support removal. A shared photograph may support selective review. Family or estate responsibility may justify notice. Direct exposure to a circulating artifact may justify a rapid challenge even when the recipient owns none of the source material. We call the procedural expression of these claims \emph{affected-party standing}. Standing does not give every claimant an equal veto; it requires the provider to receive the claim, preserve the disputed evidence, identify the current authorization basis, and explain a remedy fitted to the material and its reach.

\subsection{Fluency can conceal the edge of the person}

Morris and Brubaker describe generative ghosts as a design space in which postmortem agents can produce novel content rather than replay a fixed archive \citep{morris2025generativeghosts}. Karpus and Strasser ask when a digital replica might count as a genuine extension of a person's identity \citep{Karpus_2025}. Kozlovski and colleagues argue that interactive deadbots are marked by technological, social, philosophical, legal, and regulatory indeterminacy \citep{Kozlovski_2026}. These accounts establish that generativity changes both what the artifact can do and what kind of object it appears to be.

The case comparison adds an accountability consequence: capability also changes where evidence can be found. Dadbot's authored branches and recordings left a visible path back to John Vlahos's participation, including moments when the archive had no answer. Project December generated an expansive conversational presence from a comparatively small record. Eternos began with direct participation but placed later speech inside a commercial, model-mediated system. Technical age is secondary here. What matters is whether the system preserves the seam between a person's contribution and a model's extrapolation.

That seam matters because fluency can be a governance liability. A familiar voice and coherent first-person answer can make a thin source record feel complete. A quotation, source link, uncertainty cue, or refusal to improvise does the opposite: it keeps the edge of the archive available for inspection. Silence can therefore carry evidence. It records that the represented person did not leave an answer rather than allowing the model's confidence to occupy the gap.

Boundedness has accountability value only when it exposes provenance and limits. A fixed memorial can still use disputed material, just as a generative system can expose excellent provenance. Fidelity should therefore include the visibility of limits. Evaluations that reward resemblance and conversational continuity without testing source attribution, correction, uncertainty, and graceful non-response may select for the very fluency that makes accountability harder.

\subsection{Circulation changes who can make a claim}

Bukingolts describes mimetic AI as a creation and distribution pipeline involving targets, operators, audiences, creators, intermediaries, and regulators \citep{Bukingolts_2025}. Leuenberger argues that voice carries a distinctive relation to identity, making voice cloning more than the reproduction of sound \citep{leuenberger2025the}. Heugas traces how digitalization unsettles legal control over a person's image and likeness \citep{Heugas_2021}. Our adjacent-system stratum supplies the product layer behind those arguments: voice, likeness, and persona capabilities can be detached from the context in which they were collected and reassembled elsewhere.

The Qiao case shows why circulation changes more than audience size. A creator--subject relation became an arrangement involving relatives, fans, platforms, commercial promotion, and possible re-uploads. Each new actor changed what evidence mattered. A viewer needed a synthetic label. The family needed an authorization record and a usable removal route. A platform needed preserved evidence and a response to replication. The artifact had moved beyond the product page on which its original disclosure might have appeared.

A synthetic label can therefore disclose fabrication while leaving unauthorized circulation untouched. Family authorization, in turn, would not establish that the generated words were grounded in anything Qiao Renliang said. Provenance, authorization, and contestability answer different questions, and all three have to travel with the artifact. This is why product categories alone cannot determine governance. A private family archive and a promotional public-figure video may use similar synthesis techniques while creating different claimants, circulation risks, and evidentiary duties.

\subsection{Relational authority turns disclosure into an evidence chain}

Relational authority connects disclosures that grief technology usually presents separately. An authority record establishes the basis and scope of creation. Source boundaries distinguish recorded, retrieved, scripted, and inferred speech. A circulation checkpoint keeps authorization and provenance attached when the artifact leaves its original interface. Affected-party review links an objection to evidence, a temporary hold, a decision, and an appeal. Export, pause, transfer, and deletion then separate the possible meanings of exit. A break at any stage weakens the authority carried into the next.

One alternative explanation is that public pages are simply the wrong place to look: careful providers may reveal stronger controls during onboarding or handle disputes well in private. That possibility remains real, but it does not remove the public governance problem. Account holders may reach onboarding; represented people, relatives, co-caregivers, journalists, and regulators often cannot. Public inspectability is the only evidence surface these actors are guaranteed to share. Hidden safeguards may improve provider practice, but they cannot guide a pre-use decision or support an outsider's claim until their existence and route become visible.

The chain also creates a research program that moves beyond counting disclosures. User studies can compare recorded, retrieved, and inferred speech while measuring whether people understand the source boundary. Interviews can examine how families and co-caregivers divide authority over shared materials. Field audits can submit standardized objections, corrections, export requests, and deletion requests, then record evidence demands, interim holds, response time, and remedy. Longitudinal work can test pause and exit without treating continued use as pathology or disengagement as an inherently better outcome. Deeper case archives can track whether premortem instructions continue to constrain later models and whether controls survive provider change.

Relational authority also reaches the conduct of research. A grief narrative does not become ethically neutral when it becomes publicly accessible. It can carry family conflict, intimate messages, voices, likenesses, and representations of people who cannot consent. In this study, sensitive case material was reduced to facts needed for the analysis, intimate passages were paraphrased rather than reproduced, and the people involved were not contacted. Future case research should establish consent before combining living participants' relational records, keep quotation distinct from inference, define how contributed material can be withdrawn, and avoid republishing a synthetic artifact simply to illustrate its harm. Such decisions are part of the evidence chain itself. They apply the paper's central claim to researchers: authority should remain legible when evidence changes hands.

Relational authority makes a claim that future work can contest: accountability in AI resurrection depends less on how much a platform discloses than on whether the representation remains answerable to the people and records from which it draws legitimacy. If authority disappears each time an artifact changes models, hands, or audiences, even a perfectly labeled resurrection can remain unaccountable.

\section{Limitations and Future Work}

This corpus maps a public evidence surface rather than a global market. English-language retrieval, stable pages, and publicly documented services are overrepresented; small, discontinued, app-only, and region-specific systems are easier to miss. Archived pages and dated evidence preserve the basis of our coding, but the market will continue to change. A multilingual replication should test whether the three analytical positions travel across regions and whether different legal and cultural settings redistribute relational authority in ways this corpus cannot reveal.

Public documentation captures promises and routes visible before use; it cannot verify private onboarding, live model behavior, moderation, deletion, security, or support outcomes. The next empirical layer should turn the evidence chain into a field audit. Standardized creation, objection, correction, export, and deletion requests could test what evidence providers demand, whether disputed artifacts are paused, how long review takes, and which remedies survive model or provider change. That design would connect public inspectability to operational accountability without treating one as a proxy for the other.

One primary coder applied the current codebook. Removing 20 low-confidence systems preserved the headline direction, although that sensitivity check is not an intercoder-reliability estimate. Family assignments, thin-information thresholds, and interpretations of generic contact channels still involve judgment. The next audit should use independent coding before adjudication, report agreement by field, and publish disagreement patterns as evidence about where the concepts themselves remain unstable.

The four cases are analytic contrasts, not prevalence evidence or ethnographies. Their public records are uneven: reporting may omit private authorization, family disagreement, deletion, and later system changes, while provider materials describe public presentation rather than verified behavior. A deeper case-study program could reconstruct each chronology with direct participants, relatives, creators, providers, and preserved artifacts. It could then examine whether premortem instructions constrain later models, how relatives negotiate shared records, and what happens when an artifact moves from private remembrance into public circulation.

Lived consequences remain a separate question. The audit cannot establish dependence, renewed distress, healing, closure, or effects on children, older adults, families, and caregivers. Longitudinal research can instead ask how source boundaries, pause controls, recipient choice, and public circulation alter people's experiences over time. Comparative legal analysis can run alongside that work, testing how privacy, personality, consumer-protection, and digital-replica rules translate the same relational claims into different forms of standing and remedy.

\section{Conclusion}

Beloved afterlives expose a mismatch between platform action and relational authority. A system can begin with one person's upload, but the records, identity, and consequences do not remain with that account. Across 93 systems, creation was easier to inspect than represented-subject verification, source grounding, affected-party objection, deletion, export, or exit. Human systems showed consent becoming incomplete as representations persist. Companion-animal systems began where subject consent is unavailable and shared care must allocate authority. Adjacent systems showed how representational capacity can travel before anyone calls it an afterlife.

Four public cases supplied the movement behind those counts. Premortem participation made the origins of Dadbot and Michael Bommer's Eternos representation unusually legible, then raised questions about later models, recipients, and providers. Project December showed how intimate access can support generative extrapolation without documented authorization from the deceased. The Qiao Renliang videos carried third-party creation into public circulation, where a family's objection made the missing review route visible.

The paper's central contribution is relational authority: authorization must remain traceable as a representation changes mode, hands, providers, and audiences. That claim changes the unit of governance from a creator--product transaction to an accountability chain. Before generated speech, the chain needs an authority record and a source boundary. Before circulation, it needs authorization and provenance that travel with the artifact. When a claim is contested, it needs affected-party review and evidence preservation. At exit, it needs distinct paths for export, pause, transfer, and deletion. AI resurrection does not end when a likeness speaks. Its governance begins when authority starts to move.

\bibliographystyle{plainnat}
\bibliography{refs}

\end{document}